\documentclass[letter,11pt,DIV=12,abstract=true,numbers=noenddot,titlepage=false,twocolumn=false,draft=false]{scrartcl}
\pdfoutput=1

\makeatletter
\DeclareOldFontCommand{\rm}{\normalfont\rmfamily}{\mathrm}
\DeclareOldFontCommand{\sf}{\normalfont\sffamily}{\mathsf}
\DeclareOldFontCommand{\tt}{\normalfont\ttfamily}{\mathtt}
\DeclareOldFontCommand{\bf}{\normalfont\bfseries}{\mathbf}
\DeclareOldFontCommand{\it}{\normalfont\itshape}{\mathit}
\DeclareOldFontCommand{\sl}{\normalfont\slshape}{\@nomath\sl}

\usepackage[affil-it]{authblk}
\usepackage[numbers,sort&compress]{natbib}

\usepackage[dvips]{graphicx}

\usepackage{siunitx}
\usepackage{array,amsmath,amsthm,feynmf,mathtools}
\usepackage{autobreak}
\usepackage{multirow}
\usepackage{mathtools}
\usepackage{epsfig}
\usepackage{graphicx}
\usepackage{xcolor}
\usepackage{slashed}
\usepackage{amssymb}
\usepackage{bm}
\usepackage{bbm}
\usepackage[export]{adjustbox}
\usepackage{enumitem}
\usepackage[colorlinks=true,linkcolor=blue,citecolor=blue]{hyperref}

\allowdisplaybreaks

\begin{document}


\renewcommand\Authands{, }

\boldmath
\title{Electron EDM and $\Gamma(\mu \to e \gamma)$ in the 2HDM}
\unboldmath

\date{\today}
\author[a]{Wolfgang~Altmannshofer%
        \thanks{\texttt{waltmann@ucsc.edu}}}
\author[b]{Beno\^it~Assi%
        \thanks{\texttt{assibt@ucmail.uc.edu}}}
\author[b]{Joachim~Brod%
        \thanks{\texttt{joachim.brod@uc.edu}}}
\author[a]{Nick~Hamer%
        \thanks{\texttt{nhamer@ucsc.edu}}}
\author[c]{J.~Julio%
        \thanks{\texttt{julio@brin.go.id}}}
\author[d]{Patipan~Uttayarat%
        \thanks{\texttt{patipan@g.swu.ac.th}}}
\author[b]{Daniil~Volkov%
        \thanks{\texttt{volkovdi@mail.uc.edu}}}
	\affil[a]{{\large Department of Physics and Santa Cruz Institute for Particle Physics\\ University of California, Santa Cruz, CA 95064, USA}}
	\affil[b]{{\large Department of Physics, University of Cincinnati, Cincinnati, OH 45221, USA}}
	\affil[c]{{\large National Research and Innovation Agency, KST B. J. Habibie, South Tangerang 15314, Indonesia}}
	\affil[d]{{\large Department of Physics, Srinakharinwirot University, 114 Sukhumvit 23rd Rd., Wattana, Bangkok 10110, Thailand}}

\maketitle

\begin{abstract}
  We present the first complete two-loop calculation of the electric
  dipole moment of the electron, as well as the rates of the
  lepton-flavor violating decays $\mu \to e + \gamma$ and
  $\tau \to e/\mu + \gamma$, in the unconstrained two-Higgs doublet
  model. We include the most general Yukawa interactions of the Higgs
  doublets with the Standard Model fermions up to quadratic order, and
  allow for generic phases in the Higgs potential. A \texttt{python}
  implementation of our results is provided via a public git
  repository.
\end{abstract}

\section{Introduction}

The Standard Model of particle physics (SM) is remarkably successful
in describing the interactions and decays of all known elementary
particles, as probed mainly in collider experiments. It falls short,
however, of explaining the creation of these very particles in the
early universe: the amount of CP violation in the SM is insufficient,
and the electroweak phase transition is only of second
order~\cite{Bernreuther:2002uj, Morrissey:2012db}. This has motivated
the study of various models of baryogenesis. A popular scenario is
electroweak baryogenesis~\cite{Kuzmin:1985mm} in the context of the
two-Higgs doublet model (2HDM)~\cite{Deshpande:1977rw}, where the
missing CP violation is supplied by the complex phases associated with
the Higgs sector that is extended with respect to the SM by adding a
second Higgs doublet. It is mainly the phases in the Yukawa couplings
to the third fermion generation that play a role in models of
electroweak baryogenesis~\cite{Huber:2006ri, deVries:2017ncy,
  DeVries:2018aul, Fuchs:2020uoc}. The Higgs potential of the 2HDM
allows for a first-order phase transition for an appropriate choice of
parameters~\cite{Bernreuther:2002uj, Morrissey:2012db}.

The main constraint on these additional phases arise from the
non-observation of electric dipole moment of elementary systems such
as the electron, the neutron, or atomic and molecular
systems~\cite{Engel:2013lsa}. If one allows for the presence of phases
in all the Yukawa couplings, however, it is easy to arrange for the
bounds to essentially disappear due to the cancellation of the various
contributions. It is therefore imperative to include as many
observables as possible when testing the 2HDM. In this work, we
consider two leptonic observables, to wit, the electric dipole moment
(EDM) of the electron, and the lepton-flavor violating decays
$\mu \to e + \gamma$ and $\tau \to e/\mu + \gamma$. We calculate the
leading contributions to these observables in the most general 2HDM,
which allows for general complex, flavor non-diagonal Yukawa
couplings. Our results for the electron EDM extend the results
presented in Ref.~\cite{Altmannshofer:2020shb} by allowing for general
CP and flavor violation in the Yukawa couplings. See
Sec.~\ref{sec:conclusions} for a detailed comparison.

There are strong experimental constraints on the electron EDM
~\cite{Roussy:2022cmp} and the rare decays
$\mu \to e + \gamma$~\cite{MEG:2016leq, MEGII:2023ltw}
$\tau \to \mu + \gamma$~\cite{Belle:2021ysv}, and
$\tau \to e + \gamma$~\cite{BaBar:2009hkt}. The phenomenological
implications of these measurements are beyond the scope of this work
and will be explored in a future publication.

A brief summary of the 2HDM is contained in Sec.~\ref{sec:2hdm}. In
particular, we allow for generic phases and flavor violation in the
Higgs potential and in the Yukawa couplings. The main contributions to
the electron EDM and radiative lepton decays arise from various one-
and two-loop diagrams. We outline the calculation of these diagrams in
Sec.~\ref{sec:calc}, and present our results for the electron EDM in
Sec.~\ref{sec:edm} and for the radiative lepton decays in
Sec.~\ref{sec:leg}. In Sec.~\ref{sec:conclusions} we discuss checks on
our calculation, compare to the literature, and provide a link to
auxiliary files containing our results in computer-readable form.

\section{The unconstrained 2HDM} \label{sec:2hdm}

We start with a brief summary of the most general two-Higgs doublet
model, in order to establish our notation and conventions. We follow
the notation of Refs.~\cite{Davidson:2005cw, Haber:2006ue,
  Boto:2020wyf,Altmannshofer:2020shb}. In the unconstrained 2HDM, the
SM scalar sector is extended by an additional scalar doublet, such
that the scalar potential depends on two $SU(2)_L$ doublets $\Phi_1$
and $\Phi_2$, with hypercharge $+1/2$:
\begin{equation}\label{eq:genpot}
\begin{split}
V(\Phi_1,\,\Phi_2)
& = m_{11}^2\Phi_1^\dagger\Phi_{1}
    + m_{22}^2\Phi_2^\dagger\Phi_{2}
     - \Big(m_{12}^2\Phi_1^\dagger\Phi_{2}+\text{c.c.}\Big) \\
& \quad + {\textstyle\frac{1}{2}}\lambda_1\big(\Phi_1^\dagger\Phi_{1}\big)^2
     + {\textstyle\frac{1}{2}}\lambda_2\big(\Phi_2^\dagger\Phi_{2}\big)^2
     + \lambda_3\big(\Phi_1^\dagger\Phi_{1}\big)\big(\Phi_2^\dagger\Phi_2\big)
     + \lambda_4\big(\Phi_1^\dagger\Phi_{2}\big)\big(\Phi_2^\dagger\Phi_1\big) \\
& \quad + \Big({\textstyle\frac{1}{2}}\lambda_5\big(\Phi_1^\dagger\Phi_2\big)^2
     + \big( \lambda_6 \Phi_1^\dagger\Phi_{1} +
             \lambda_7 \Phi_2^\dagger\Phi_{2} \big)
       \big(\Phi_1^\dagger\Phi_2\big) + \text{c.c.}\Big)\,.
\end{split}
\end{equation}
The parameters $m_{11}^2$, $m_{22}^2$ and
$\lambda_1,\,\ldots, \lambda_4$ are real parameters, while $m_{12}^2$,
$\lambda_5$, $\lambda_6$, $\lambda_7$ may be complex. In general, both
doublets may acquire a vacuum expectation value. Assuming that the
parameters in the potential are chosen such that the vacuum respects
electromagnetic gauge symmetry, the vacuum expectation values of the
two doublets take the form
\begin{equation}
\langle\Phi_1\rangle = \frac{1}{\sqrt{2}}\begin{pmatrix}0\\ v_1\end{pmatrix}\,,\qquad
\langle\Phi_2\rangle = \frac{1}{\sqrt{2}}\begin{pmatrix}0\\ v_2 e^{i\zeta}\end{pmatrix}\,,
\end{equation}
where $v^2\equiv v_1^2 + v_2^2 = (246\text{ GeV})^2$, and $\zeta$ is a
possible relative phase between the vacuum expectation values. The
size of $v_1$ and $v_2$ are determined by the parameters of the Higgs
potential. The relevant expressions are given in
App.~\ref{app:higgs_potential}. Rephasing invariance can be used to
set $\zeta=0$; we follow this convention throughout the paper.

To isolate the physical degrees of freedom in the scalar sector, it is
convenient to transform the fields into the so-called Higgs basis,
\begin{equation}
\begin{pmatrix}
\Phi_1 \\ \Phi_2
\end{pmatrix}=
\begin{pmatrix}
\cos\beta & -\sin\beta\\
\sin\beta & \cos\beta
\end{pmatrix}
\begin{pmatrix}
H_1\\
H_2
\end{pmatrix}\,,
\end{equation}
with $\tan\beta = v_2/v_1$, such that only the field $H_1$ receives a
vacuum expectation value. In the Higgs basis, the potential reads
\begin{equation}\label{eq:HBpot}
\begin{split}
\mathcal{V}(H_1,\,H_2)
& = Y_1 H_1^\dagger H_{1}
    + Y_2 H_2^\dagger H_{2}
    + \Big(Y_3  H_1^\dagger H_{2} +\text{c.c.}\Big) \\
& \quad + {\textstyle\frac{1}{2}}Z_1\big( H_1^\dagger H_{1}\big)^2
    + {\textstyle\frac{1}{2}}Z_2\big( H_2^\dagger H_{2}\big)^2
    + Z_3\big( H_1^\dagger H_{1}\big)\big( H_2^\dagger H_2\big)
    + Z_4\big( H_1^\dagger H_{2}\big)\big( H_2^\dagger H_1\big) \\
& \quad + \Big({\textstyle\frac{1}{2}}Z_5  \big( H_1^\dagger H_2\big)^2
    + \big(Z_6\, H_1^\dagger H_1 + Z_7\, H_2^\dagger H_2 \big)
      \big(H_1^\dag H_2\big) +\text{c.c.}\Big) \,,
\end{split}
\end{equation}
with new parameters $Y_i$, $Z_i$ that are linear combinations of the
original parameters $m_{ij}^2$ and $\lambda_i$ -- see
App.~\ref{app:higgs_potential}. In the Higgs basis, the components of
the scalar fields can be written as
\begin{equation}
H_1 =
\begin{pmatrix}
G^+\\
\frac{1}{\sqrt{2}}\big(v + \varphi_1^0 + i G^0\big)
\end{pmatrix}\,, \qquad 
H_2 =
\begin{pmatrix}
H^+ \\
\frac{1}{\sqrt{2}}\big(\varphi_2^0 + i a^0\big)
\end{pmatrix}\,.
\end{equation}
The unphysical ``would-be'' Goldstone boson fields are denoted by
$G^+$ and $G^0$. The physical charged Higgs bosons $H^\pm$ have a
squared mass given by
\begin{equation}
 M_{H^\pm}^2 = Y_2 + \frac{1}{2}Z_3 v^2 \,.
\end{equation}
As we allow for $CP$ violation, the neutral $CP$-even Higgs bosons
$\varphi_1^0$ and $\varphi_2^0$ and the neutral $CP$-odd Higgs boson
$a^0$ mix, with the corresponding symmetric squared-mass matrix
$\mathcal{M}^2$ given by
\begin{equation}\label{eq:neutralMassMtx}
 \frac{\mathcal{M}^2}{v^2} =
\begin{pmatrix}
Z_1 & \text{Re}(Z_6) & -\text{Im}(Z_6) \\[0.5mm]
    & Y_2/v^2+\frac{1}{2}Z^+_{345} & -\frac{1}{2}\text{Im}(Z_5 ) \\[1mm]
    & &  Y_2/v^2+\frac{1}{2}Z^-_{345}
\end{pmatrix}\,,
\end{equation}
where $Z_{345}^\pm = Z_3 + Z_4 \pm \text{Re}(Z_5)$. This symmetric
mass matrix is diagonalized by a special orthogonal transformation
that can be parameterized as
\begin{equation}
\label{eq:HBtoMassRot}
\begin{pmatrix}
h_1 \\ h_2 \\ h_3
\end{pmatrix} = \begin{pmatrix}
q_{11} & \text{Re}(q_{12}) & \text{Im}(q_{12})\\
q_{21} & \text{Re}(q_{22}) & \text{Im}(q_{22})\\
q_{31} & \text{Re}(q_{32}) & \text{Im}(q_{32})
\end{pmatrix}
\begin{pmatrix}
\varphi_1^0 \\
\varphi_2^0 \\
a^0
\end{pmatrix}\,.
\end{equation}
We denote the squared masses of the three neutral mass eigenstate
Higgs bosons $h_k$ with $M_{h_k}^2$, for $k=1,2,3$.

The elements of the rotation matrix in Eq.~\eqref{eq:HBtoMassRot} can
be parametrized by three angles
\begin{equation}
\begin{pmatrix}
q_{11} & \text{Re}(q_{12}) & \text{Im}(q_{12})\\
q_{21} & \text{Re}(q_{22}) & \text{Im}(q_{22})\\
q_{31} & \text{Re}(q_{32}) & \text{Im}(q_{32})
\end{pmatrix}
=
\begin{pmatrix}
c_{12} c_{13}&- s_{12} c_{23} - c_{12} s_{13} s_{23}& s_{12} s_{23}-c_{12}s_{13}c_{23}\\
s_{12} c_{13}& c_{12} c_{23} - s_{12} s_{13} s_{23} &-c_{12} s_{23}-s_{12}s_{13}c_{23}\\
s_{13} & c_{13} s_{23} & c_{13} c_{23}
\end{pmatrix}\,,
\end{equation}
where $s_{ij}$ ($c_{ij}$) is $\sin\theta_{ij}$
($\cos\theta_{ij}$). This parameterization is used in the
\texttt{python} code, see Sec.~\ref{sec:conclusions}.

Next, we introduce the Yukawa interactions of the doublets $\Phi_1$
and $\Phi_2$ with the SM fermions. In the gauge eigenstate basis, the
most general interaction Lagrangian can be written as
\begin{equation}\label{eq:Yukawa:Gauge}
\mathcal{L}_\text{Yuk}
 = - \sum_{a = 1}^2 \sum_{ij=1}^3 \Big(
                   \bar Q_{L,i} \hat{Y}_{d,ij}^{(a)} d_{R,j} \Phi_a
                 + \bar Q_{L,i} \hat{Y}_{u,ij}^{(a)} u_{R,j} \tilde\Phi_a
                 + \bar L_{L,i} \hat{Y}_{\ell,ij}^{(a)} \ell_{R,j} \Phi_a
                \Big) + \text{c.c.} \,,
\end{equation}
where $Q_L$, $L_L$ denote the doublets of left-handed quark and lepton
fields, $d_R$, $u_R$, and $\ell_R$ the right-handed up, down, and
lepton singlet fields, and $\tilde\Phi_a = i \sigma_2 \Phi_a^*$ are
the charge-conjugated Higgs fields. The Yukawa couplings
$\hat{Y}^{(a)}_f$, with $f=u,d,\ell$, are general complex $3\times3$
matrices, and $i,j=1,2,3$ are flavor indices.

Not all entries of the Yukawa matrices are physical. The physical
content can be displayed more succinctly by expressing the Yukawa
Lagrangian $\mathcal{L}_\text{Yuk}$ in terms of Higgs and fermion mass
eigenstates. Following the notation of Ref.~\cite{Boto:2020wyf} we
have
\begin{equation}
\begin{split}
\mathcal{L}_\text{Yuk}
 & \supset
  - \sum_k \sum_{ij} h_k \bigg[
  \bar{u}_{L,i}
  \left( \frac{m_{u_i}}{v} \delta_{ij} q_{k1} + \rho_{u,ij} q^*_{k2} \right) u_{R,j} \\
& \hspace{6em} + \sum_{f=d,e}
  \bar{f}_{L,i}
  \left( \frac{m_{f_i}}{v} \delta_{ij} q_{k1} + \rho_{f,ij}^\dagger q_{k2} \right) f_{R,j}
  \bigg] + \text{c.c.}  \\
& \quad 
- \sqrt{2} H^+ \big[ \bar{u}_{L,i} \big(V \rho_d^\dagger \big)_{ij} d_{R,j}
                    - \bar{u}_{R,i} \big(\rho_u^\dagger V \big)_{ij} d_{L,j} \big]
- \sqrt{2} H^+ \bar{\nu}_{L,i} \rho_{\ell,ij}^\dagger e_{R,j} + \text{c.c.} \,,
\end{split}
\end{equation}
where we did not display the interactions of the unphysical Goldstone
bosons. The $3\times 3$ matrices ${\rho}_f$ contain general complex
entries that lead to flavor and $CP$ violation. Their relation to the
Yukawa matrices in Eq.~\eqref{eq:Yukawa:Gauge} is given by
\begin{equation}
\rho_u = \big( Y_u^{(2)} c_\beta - Y_u^{(1)} s_\beta \big) / \sqrt{2} \,, \qquad 
\rho_{d,e}^\dagger = \big( Y_{d,e}^{(2)} c_\beta - Y_{d,e}^{(1)} s_\beta \big) / \sqrt{2} \,,
\end{equation}
while the fermion masses are given by the diagonal combination
\begin{equation}
 m_{f_i} = \frac{v}{\sqrt{2}} \big({Y}_{f}^{(1)} c_\beta + Y_{f}^{(2)} s_\beta \big)_{ii} \,.
\end{equation}
The matrices $Y_f^{(a)}$ in these relations are the Yukawa couplings
in Eq.~\eqref{eq:Yukawa:Gauge}, rotated into the fermion mass
eigenstate basis. The charged Higgs interactions with quarks contain
the Cabibbo-Kobayashi-Maskawa (CKM) quark mixing matrix
$V$.\footnote{Two popular versions of the 2HDM can be obtained as the
  following limiting cases. Type-I 2HDM: $Y_f^{(1)} = 0$. In this
  case, $\rho_f = (m_f/v) \cot\beta$. Type-II 2HDM:
  $Y_{d,\ell}^{(2)} = Y_u^{(1)} = 0$. In this case,
  $\rho_u = (m_u/v) \cot\beta$, and
  $\rho_{d,\ell} = -(m_{d,\ell}/v) \tan\beta$. In particular, the
  $\rho$ matrices are real and diagonal in both
  cases~\cite{Glashow:1976nt}.}

\section{Outline of the calculation} \label{sec:calc}

In this work, we calculate predictions for low-energy observables,
namely, the electron EDM and flavor-violating radiative lepton decay
rates.\footnote{Note that the most sensitive experimental searches for
  the electron EDM are performed using molecular ions. Such searches
  are not only sensitive to the electron EDM but also to CP-violating
  electron-nucleon interactions. At the quark level, these
  interactions correspond to four-fermion contact operators which can,
  in principle, be induced at the tree-level through neutral Higgs
  exchange. Apart from regions of parameter space with large values of
  $\tan\beta$, such contributions to molecular EDMs are expected to be
  subdominant. A detailed study of the tree-level effects is beyond
  the scope of this paper and is left for future work.} The energy
scales characterizing these observables are of the order of the mass
or the decaying lepton in the latter case, or far below the electron
mass for the electron EDM.

\begin{figure}[t]
  \centering
  \includegraphics[width=12em]{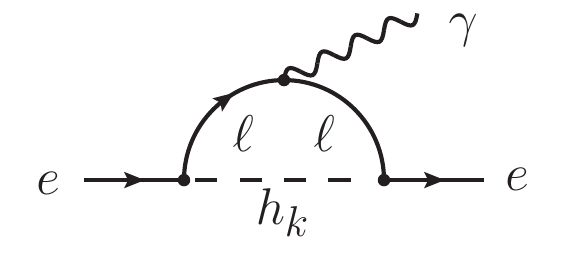}~~~~
  \includegraphics[width=12em]{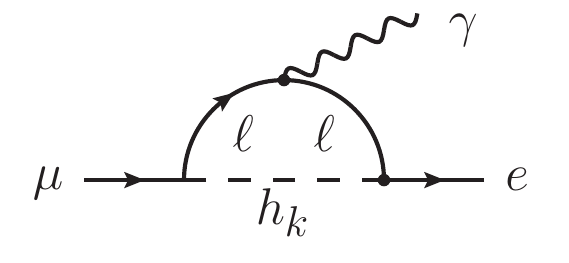}~~~~
  \includegraphics[width=12em]{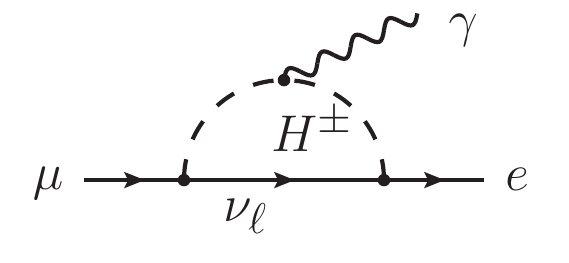}
  \caption{One-loop contributions to the EDM and anomalous magnetic
    moment of the electron (left panel), and to $\mu \to e \gamma$
    (center and right panels). Solid lines denote charged leptons
    (labelled by $\ell=e,\mu,\tau$) and the corresponding neutrinos,
    and dashed lines denote Higgs bosons (labelled by $h_k$ and
    $H^\pm$). Diagrams with charged Higgs bosons do not contribute to
    the electron EDM at one loop. \label{fig:edm:1loop}}
\end{figure}

Both observables are induced by amplitudes that are non-zero only at
one-loop order or higher (see Fig.~\ref{fig:edm:1loop}). The leading
contributions in the 2HDM arise from flavor-violating Yukawa couplings
as well as phases in the Yukawa couplings and the Higgs potential (all
other contributions are due to the CKM matrix and are as suppressed as
in the SM~\cite{Pospelov:2005pr, Yamaguchi:2020eub, Yamaguchi:2020dsy,
  Ema:2022yra}). Excluding exotic scenarios, we will assume that all
Higgs bosons in the 2HDM have masses at the electroweak scale or
above. It is, therefore, most natural to perform the calculation by
integrating out the heavy Higgs bosons, together with the top quark
and the heavy gauge bosons, and match to an effective leptonic
Lagrangian defined below the electroweak scale. (In the case that all
Higgs bosons have masses well above the weak scale, an alternative
approach would be to match to SMEFT in intermediate
steps~\cite{Davidson:2016utf, Panico:2018hal, Brod:2022bww}.)

In this work, we focus on purely leptonic observables, so there are no
appreciable effects of the strong interaction.\footnote{There is one
  notable exception to this statement: the two-loop bottom- and
  charm-quark contribution to the electron EDM receives large QCD
  corrections. However, as discussed in Ref.~\cite{Brod:2023wsh}, a
  reasonable approximation to the quite involved renormalization-group
  analysis (including next-to-leading-logarithmic QCD corrections) is
  obtained by evaluating the bottom- and charm-quark masses at the
  weak scale. We will adopt this shortcut here. \label{foot:rg}}
Furthermore, we neglect the tiny effects of QED running (they are
suppressed by powers of the fine-structure constant
$\alpha \approx 0.008$). Thus, the leading effects are captured to an
excellent approximation by a weak-scale matching calculation.

All Feynman diagrams have been calculated using the package
\texttt{MaRTIn}~\cite{Brod:2024zaz}, based on
\texttt{FORM}~\cite{Vermaseren:2000nd} and implementing the algorithms
developed in Refs.~\cite{Davydychev:1992mt, Bobeth:1999mk}. The
diagrams have been generated using
\texttt{qgraf}~\cite{Nogueira:1991ex}. The explicit form of the
Feynman rules has been obtained using the \texttt{FeynRules}
package~\cite{Alloul:2013bka}. We have also the general results in
Ref.~\cite{Brod:2019bro} to calculate some of the Feynman rules for
the unphysical Goldstone bosons that arise as virtual particles in the
diagrams. Throughout, we work in a generalized $R_\xi$ gauge for the
photon and the weak gauge bosons, and verified that the gauge
parameter drops out of all physical results. We have implemented the
background field gauge in the 2HDM following the procedure in
Ref.~\cite{Denner:1994xt}.

We choose to perform the calculation using the background field gauge
method in the electroweak sector, since this leads to a drastic
reduction in the complexity of the
calculation~\cite{Altmannshofer:2020shb}. Our implementation follows
Ref.~\cite{Denner:1994xt}. In the unbroken phase, we split all gauge
fields into quantum fields and background fields (the latter denoted
by a hat). We also perform a similar splitting for the field $H_1$ in
the Higgs basis, i.e. we have
\begin{equation}
  \hat H_1 =
\begin{pmatrix}
  \hat G^+\\
  \frac{1}{\sqrt{2}} \big( v + \hat \phi_1^0 + i \hat G^0 \big)
\end{pmatrix}\,, \qquad
  H_1 =
\begin{pmatrix}
  G^+\\
  \frac{1}{\sqrt{2}} \big( \phi_1^0 + iG^0 \big)
\end{pmatrix}\,.
\end{equation}
We then add the following gauge-fixing term to the Lagrangian,
\begin{equation}
\begin{split}
  {\mathcal L}_\text{gf}
& = - \frac{1}{2\xi}
      \bigg[ \big( \delta^{ac} \partial_\mu
                   + g_2 \epsilon^{abc} \hat W_\mu^b \big) W^{c,\mu}
             -ig_2 \xi \frac{1}{2}
             \big( \hat H_{1,i}^\dagger \sigma_{ij}^a H_{1,j} 
                   - H_{1,i}^\dagger \sigma_{ij}^a \hat H_{1,j}\big) \bigg]^2 \\
& \quad - \frac{1}{2\xi}
          \bigg[ \partial_\mu B^\mu 
                 +ig_1 \xi \frac{1}{2}
                 \big( \hat H_{1,i}^\dagger H_{1,i} 
                       - H_{1,i}^\dagger \hat H_{1,i}\big) \bigg]^2 \,.
\end{split}
\end{equation}
Here, $g_1$ and $g_2$ are the gauge couplings associated with the hypercharge
$U(1)_Y$ boson $B_\mu$ and the
triplet of weak $SU(2)_L$ gauge bosons $W_\mu^a$, respectively, and $\xi$ is an arbitrary gauge
parameter, chosen to be the same for all weak gauge bosons, to avoid
tree-level mixing between the photon and the $Z$ boson (see
Ref.~\cite{Denner:1994xt} for details). Moreover,
$s_w \equiv \sin\theta_w$ denotes the sine of the weak mixing angle,
and $c_w \equiv \cos\theta_w$. The Feynman rules are obtained by
expressing all fields in the broken phase and rotating to mass
eigenstates. The ghost Lagrangian is constructed in the usual
way~\cite{Denner:1994xt}. For our calculation, only the Feynman rules
with one background photon field are needed. In our conventions,
$\hat A^\mu = c_w \hat B^\mu - s_w \hat W^{3,\mu}$.

\section{Contributions to the electron EDM} \label{sec:edm}

The effective Lagrangian below the electroweak scale can be defined as
\begin{equation} \label{eq:L:eff}
\mathcal L_\text{eff}
 = - \frac{d_e}{2} (\bar e \sigma^{\mu\nu} i\gamma_5 e) F_{\mu\nu} \,.
\end{equation}
Here, $\sigma^{\mu\nu} = \tfrac{i}{2} [\gamma^\mu, \gamma^\nu]$, and
$F_{\mu\nu} = \partial_\mu A_\nu - \partial_\nu A_\mu$. We follow the
conventions in Ref.~\cite{Denner:1991kt} which imply that the
covariant derivative acting on lepton fields is given by
\begin{equation}
  D_\mu = \partial_\mu + ieQ_q A_\mu \,.
\end{equation}
As we are neglecting any effects of QED running, the effective
Lagrangian~\eqref{eq:L:eff} can be thought of as equally valid at
scales of the order of the electron mass. We split the contribution to
the dipole coefficient according to the number of loops, i.e.
\begin{equation}
  d_e = d_e^\text{one-loop} + d_e^\text{two-loop} + \ldots \,.
\end{equation}
Contributions at three loops or higher are indicated by the ellipsis
and are not considered in this work.

At one-loop, only the diagrams with neutral Higgs exchange contribute
(see Fig.~\ref{fig:edm:1loop}, left panel). We find
\begin{equation}\label{eq:edm:1loop}
\begin{split}
  d_e^\text{one-loop}
 & = \quad \frac{e}{32\pi^2} \sum_{i=1}^3 \sum_{k=1}^3 \frac{m_{\ell_i}}{M_{h_k}^2}
           \text{Im} \big\{\rho_{\ell,i1}^* \rho_{\ell,1i}^* q_{k2}^2 \big\}
           \big( 3 + 2 \log x_{\ell_i h_k} \big) \\
 & \quad + \frac{e}{16\pi^2} \sum_{k=1}^3 \frac{m_{e}^2}{M_{h_k}^2 v}
           q_{k1} \text{Im} \big\{\rho_{\ell,11}^* q_{k2} \big\}
           \big( 3 + 2 \log x_{\ell_1 h_k} \big) \,,
\end{split}
\end{equation}
where here and below we define the mass ratios
$x_{a\, b} = M_a^2/M_b^2$. This result is obtained by performing the
matching to leading-logarithmic and next-to-leading logarithmic order,
expanding up to second order in the external momenta, but keeping only
a linear power in the electron mass. The Dirac equation has been used
to eliminate contributions that vanish via the electron equations of
motion. Note that the result of the one-loop matching is finite only
after combining it with the one-loop matrix element calculated in the
effective theory. Any dependence on the renormalization scheme
employed in the effective theory similarly cancels in this
combination.

We pause here to make an important comment about the presentation of
our results. It is important to realize that the mixing angles
$q_{k1}$ and $q_{k2}$, $k=1,2,3$, are not all independent, but
constrained by various relations arising from the orthogonality of the
transformation matrix in Eq.~\eqref{eq:HBtoMassRot}. For instance,
using
$\sum_k q_{k1} \text{Im}(q_{k2}) = \sum_k q_{k1} \text{Re}(q_{k2}) =
0$, it is easy to see that the three terms in the second line in
Eq.~\eqref{eq:edm:1loop} reduce to two terms that are each simply
proportional to $\log(x_{h_1 h_k})$, for $k=2,3$. The first line,
however, would look quite awkward after imposing these relations
explicitly. We will therefore, in general, not explicitly impose the
orthogonality conditions on our results, unless they are required to
render the result finite and / or gauge independent (see, for
instance, Eqs.~\eqref{eq:edm:hZ} and~\eqref{eq:eEDM:bosonic}). It is
important to keep in mind that all spurious mixing angles should be
eliminated before using our results in phenomenological applications.

\begin{figure}[t]
  \centering
  \includegraphics[width=12em]{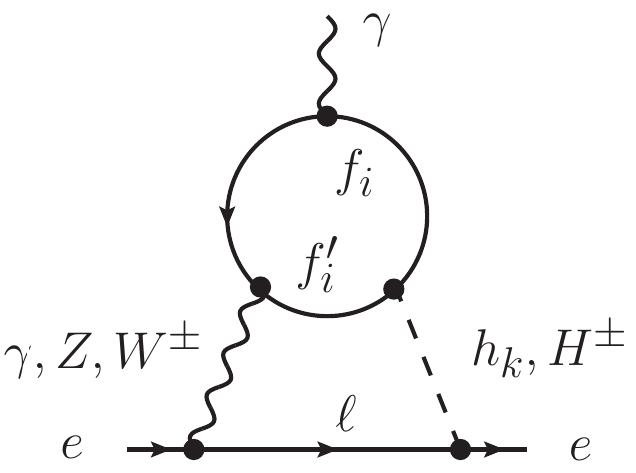}~~~
  \includegraphics[width=12em]{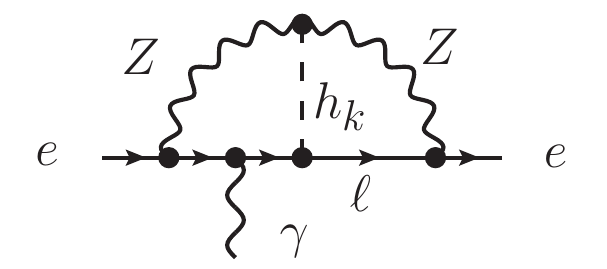}
  \caption{Representative two-loop diagrams contributing to the
    electron EDM. \label{fig:BZ:fermion}}
\end{figure}

It has been pointed out long ago by Weinberg~\cite{Weinberg:1989dx}
and Dicus~\cite{Dicus:1989va}, and by Barr and Zee~\cite{Barr:1990vd}
that numerically large contributions to the dipole operators might
arise from two-loop diagrams if there is a large hierarchy between the
various Yukawa couplings. Accordingly, we calculate all {\em leading}
two-loop contributions that are quadratic in the Yukawa couplings. (In
particular, we do not keep two-loop contributions that probe
combinations of couplings that are already present at one loop --
these would constitute small radiative corrections to the leading
one-loop results.) It is convenient to split the two-loop contribution
into several terms as follows
\begin{equation}
  d_e^\text{two-loop}
= \sum_{f_i = t,b,c,\tau} d_e^{f_i h}
  + \sum_{f_i = t,c,\tau} d_e^{f_i H^\pm}
  + \big( d_e^{hZ} + d_e^\text{kin.} \big)
  + \big( d_e^{H^\pm \gamma} + d_e^{H^\pm Z} + d_e^{H^\pm W} \big) \,.
\end{equation}
Here, the $d_e^{f_i h}$ and $d_e^{f_i H^\pm}$ represent fermionic
Barr-Zee-type contributions with exchanges of virtual neutral and
charged Higgs bosons, respectively, while the contributions within the
first and second parentheses arise from diagrams that involve vertices
from the Higgs kinetic terms and the Higgs potential,
respectively. All the individual terms in this sum are separately
gauge invariant. We discuss these terms in the following and present
the explicit results.

The contributions of Barr-Zee diagrams with internal fermion loops and
neutral Higgs bosons (see Fig.~\ref{fig:BZ:fermion}, left panel) are
given by
\begin{equation}
\begin{split}\label{eq:res:defi}
  & d_e^{f_i h}
  = \frac{N_c(f_i)e\alpha Q_{f_i}}{16\pi^3}
    \sum_{k=1}^3 \frac{m_{f_i}}{M_{h_k}^2} \\
& \times \bigg\{ \mp
  \bigg[ Q_{f_i} f_{1}(x_{f_i h_k})
         + \frac{v_{f_i}^Z v_e^Z}{4} f_{3}(x_{f_i h_k}, x_{f_i Z}) \bigg]
  \text{Im}( \rho_{f,ii}^* q_{k2} )
  \Big( \frac{m_{e}}{v} q_{k1} + \text{Re} (\rho_{\ell,11}^* q_{k2}) \Big) \\
& \qquad + \bigg[ Q_{f_i} f_{2}(x_{f_i h_k})
                 + \frac{v_{f_i}^Z v_e^Z}{4} f_{4}(x_{f_i h_k}, x_{f_i Z}) \bigg]
  \text{Im}( \rho_{\ell,11}^* q_{k2} )
  \Big( \frac{m_{f_i}}{v} q_{k1} + \text{Re} (\rho_{f,ii}^* q_{k2}) \Big)
  \bigg\} \,,
\end{split}
\end{equation}
where the upper sign is for up-type fermions ($f_i = t,c$; weak
isospin $T_3^{f_i} = 1/2$) and the lower sign for down-type fermions
($f_i = b,\tau$; weak isospin $T_3^{f_i} = -1/2$). The index $i=3(2)$
for third-(second-)generation fermions. Moreover,
$v_f^Z \equiv (T_3^f - 2Q_f s_w^2)/s_wc_w$ is the vectorial $Z$
coupling to fermion $f$. $Q_f$ is the charge of fermion $f$ in units
of the positron charge $e$, and $N_c(t) = N_c(b) = N_c(c) = 3$ and
$N_c(\tau) = 1$ are color factors. In these results, we have
integrated out the bottom and charm quarks as well as the tau lepton
together with the weak-scale particles, in order to avoid the
unnecessarily complicated renormalization-group analysis mentioned in
footnote~\ref{foot:rg}. The bulk of the leading QCD corrections can
(and should be) taken into account by evaluating the bottom and charm
masses at the electroweak scale. Note that the contributions of
virtual up, down, and strange quarks as well as electrons and muons
are strongly suppressed by their small masses, and are not included
here.\footnote{Constraints on $CP$-odd Yukawa couplings of the Higgs
  bosons to the three light quarks receive much stronger constraints
  from hadronic EDMs. They will be presented in a forthcoming
  publication. See also Ref.~\cite{Brod:2022bww}.} Note also that the
contributions with virtual $Z$-boson exchange are suppressed by the
small value of $v_e^Z$ and have been included here mainly for
completeness. The dimensionless two-loop functions appearing in this
result are well-known~\cite{Brod:2022bww}; they are given explicitly
by
\begin{align}
  f_{1}(x) & = \Phi\bigg(\frac{1}{4x}\bigg) \,, \label{eq:f21}
& f_{2}(x) & = (1-2x) f_{1}(x) + 2 \big(2 + \log(x) \big) \,, \\
  f_{3}(x,y) & = \frac{y}{x - y} \bigg[ f_{1}(x) - f_{1}(y) \bigg] \,,
& f_{4}(x,y) & = \frac{y}{x - y}
                  \bigg[ f_{2}(x) - f_{2}(y) \bigg] \,. \label{eq:f24}
\end{align}
The functions $\Phi(x)$ and $\Phi(x,y)$ that appear in these
expressions are defined in Ref.~\cite{Davydychev:1992mt}. The
dilogarithm is defined as
\begin{equation}
  \text{Li}_2 (x) = -\int_0^x du \, \log (1-u)/u \,.
\end{equation}
For internal fermions other than the top quark, it is an excellent
approximation to use the asymptotic values of the loop functions for
$x \ll 1$ and $y \ll 1$, given by
\begin{align} 
  f_{1}(x) & \simeq \log^2 x + \frac{\pi^2}{3} \,, \label{eq:limit:f21}
& f_{2}(x) & \simeq \log^2 x + 2 \log x + 4 + \frac{\pi^2}{3} \,, \\
  f_{3}(x,y) & \simeq \frac{y}{x - y} \big[ \log^2 x - \log^2 y \big] \,,
& f_{4}(x,y) & \simeq \frac{y}{x - y} \bigg[ \log^2 x - \log^2 y + 2 \log \frac{x}{y} \bigg] \,.
  \label{eq:limit:f24}
\end{align}

The contribution of diagrams with internal fermions and charged Higgs
bosons (see Fig.~\ref{fig:BZ:fermion}, left panel) is given, for
internal top and bottom quarks:\footnote{This result is valid for a
  unit CKM matrix. When using the actual CKM matrix, the first term in
  Eq.~\eqref{eq:eEDM:tchH} would remain unchanged, while the second
  would receive additional contributions proportional to off-diagonal
  elements of both $\rho_d$ and the CKM matrix. We note that
  down-quark EDMs would receive contributions from bosonic two-loop
  diagrams that are proportional to the same off-diagonal elements of
  $\rho_d$, but are not suppressed by small quark masses and CKM
  angles. We will consider hadronic EDMs in a future publication, and
  drop any such suppressed contributions in our results for the
  electron EDM.}
\begin{equation}\label{eq:eEDM:tchH}
\begin{split}
  d_e^{tH^\pm}
= \frac{e\alpha}{64\pi^3s_w^2} \frac{m_t}{M_{H^\pm}^2}
  \bigg( & \text{Im} \big( \rho_{u,33} \rho_{\ell,11}^* \big) f_{5}(x_{t H^\pm}, x_{t W}) \\
& \quad + \frac{m_b}{m_t} \text{Im} \big( \rho_{d,33} \rho_{\ell,11}^* \big)
          f_{6}(x_{t H^\pm}, x_{t W})\bigg) \,.
\end{split}
\end{equation}
Here, the explicit expressions for the two-loop functions are
\begin{align}
  \tilde f_{5}(x) & = 3 x - \frac{1}{2}(13-6x)\log x
                  + (2-x)(2-3x)\bigg[ \text{Li}_2(1-1/x) - \frac{\pi^2}{6} \bigg] \,, \\
  f_{5}(x,y) & = \frac{y}{x - y} \bigg[ \tilde f_{5}(x) - \tilde f_{5}(y) \bigg] \,,
  \label{eq:f25} \\
  \tilde f_{6}(x) & = -3 x + \frac{1}{2}(1-6x)\log x
                  +x(2-3x)\bigg[ \text{Li}_2(1-1/x) - \frac{\pi^2}{6} \bigg] \,, \\
  f_{6}(x,y) & = \frac{y}{x - y} \bigg[ \tilde f_{6}(x) - \tilde f_{6}(y) \bigg]
  \label{eq:f26} \,.
\end{align}

The contribution of diagrams with internal charm quarks is
\begin{equation}\label{eq:decH}
\begin{split}
  d_e^{cH^\pm}
= & \frac{e\alpha}{128\pi^3s_w^2} \frac{m_c}{M_{H^\pm}^2}
    \text{Im} \big( \rho_{u,22} \rho_{\ell,11}^* \big)
    \bigg[ 13 + 4 \log\left(\frac{m_c^2}{M_{H^\pm}^2}\right)
              + 4\log \left(\frac{m_c^2}{M_W^2}\right) \bigg]
    \frac{\log x_{W H^\pm}}{1 - x_{W H^\pm}} \,.
\end{split}
\end{equation}
This result is obtained by taking the asymptotic limit of the loop
function $f_{5}(x,y)$ for small arguments.  The contribution of
diagrams with an internal tau leptons is
\begin{equation}
  d_e^{\tau H^\pm} = \frac{e\alpha}{128\pi^3s_w^2} \frac{m_{\tau}}{M_{H\pm}^2}
             \text{Im} \big( \rho_{\ell,33} \rho_{\ell,11}^* \big)
             \frac{1}{1 - x_{W H^\pm}} \log x_{W H^\pm} \,.
\end{equation}
Here, we have again retained only the limiting value of the two-loop
function. It is interesting to point out that, in contrast to
Eq.~\eqref{eq:decH}, no large logarithm appears in this result.

A technical comment is in order. The calculation of the diagrams with
internal closed fermion loops (Fig.~\ref{fig:BZ:fermion}, left panel)
involves the evaluation of fermion traces containing one or more
powers of $\gamma_5$. For reasons of algebraic consistency, we employ
the 't~Hooft-Veltman scheme for $\gamma_5$, involving mixed
commutation and anticommutation relations (see
Ref.~\cite{tHooft:1972tcz, Breitenlohner:1977hr, Collins:1984xc} for
details). Our calculations are consistent with the available results
in the literature, after the proper inclusion of finite counterterm
contributions.

Next, we discuss the contributions from diagrams involving vertices
arising from the kinetic terms of the Higgs bosons. There are two
classes of diagrams that are separately gauge invariant. The first
class consists of the diagrams with internal $Z$ and neutral Higgs
bosons that are not of the Barr-Zee type (see
Fig.~\ref{fig:BZ:fermion}, right panel):
\begin{equation}\label{eq:edm:hZ}
\begin{split}
  d_e^{hZ} & = - \frac{\alpha^2 \big(v_e^Z\big)^2}{96\pi^2s_w c_w} \frac{1}{M_Z}
             \sum_{k=2}^3 q_{k1} \text{Im} \big( \rho_{\ell,11}^* q_{k2} \big)
             \big[ f_{7+}(x_{Z h_k}) - f_{7+}(x_{Z h_1}) \big] \\
          & \quad - \frac{\alpha^2 \big(a_e^Z\big)^2}{96\pi^2s_w c_w} \frac{1}{M_Z}
             \sum_{k=2}^3 q_{k1} \text{Im} \big( \rho_{\ell,11}^* q_{k2} \big)
             \big[ f_{7-}(x_{Z h_k}) - f_{7-}(x_{Z h_1}) \big] \,,
\end{split}
\end{equation}
where $a_f^Z \equiv - T_3^f/(s_wc_w)$ is the axial $Z$ coupling to
fermion $f$. The loop functions are
\begin{equation} \label{eq:f27pm}
\begin{split}
  f_{7\pm}(x) & = \frac{16 x^4 + 4 x^3 - (2\pm 24)x^2 \pm (18 x - 3)}{4x^3} \Phi\bigg(\frac{1}{4x}\bigg)
                 + \frac{8 x^2 - 2 x \mp 3}{2x} \\
              & \quad - \frac{4 x^5 + 3 x^4 - (1\pm 3) x^2 \pm (12 x - 3)}{x^3} \, \text{Li}_2 (1 - x)
                      - \frac{8 x^5 + 6 x^4 \mp (12 x - 3)}{12 x^3} \pi^2 \\
              & \quad - \log^2(x) \frac{8 x^5 + 6 x^4 - (2\pm 6)x^2 \pm (12 x - 3)}{4x^3}
                      + \log(x) \frac{4 x^2 + x \pm 3}{x} \,.
\end{split}
\end{equation}

All remaining diagrams with vertices from the Higgs kinetic terms give
\begin{equation}\label{eq:eEDM:bosonic}
\begin{split}
  d_e^{\text{kin.}} & = \frac{\alpha^2 v_e^Z}{64\pi^2} \frac{1}{M_Z}
             \sum_{k=2}^3 q_{k1} \text{Im} \big( \rho_{\ell,11}^* q_{k2} \big)
             \big[ f_{8a}(x_{Z h_k}, c_w^2) - f_{8a}(x_{Z h_1}, c_w^2) \big] \\
                                & \quad + \frac{\alpha^2 v_e^Z}{64\pi^2 s_w^2} \frac{1}{M_Z}
             \sum_{k=2}^3 q_{k1} \text{Im} \big( \rho_{\ell,11}^* q_{k2} \big)
             \big[ f_{8b}(x_{Z h_k}, c_w^2) - f_{8b}(x_{Z h_1}, c_w^2) \big] \\
                                & \quad + \frac{\alpha^2}{32s_w\pi^2} \frac{1}{M_W}
             \sum_{k=2}^3 q_{k1} \text{Im} \big( \rho_{\ell,11}^* q_{k2} \big)
             \big[ f_{8c}(x_{W h_k}) - f_{8c}(x_{W h_1}) \big] \\
                                & \quad + \frac{\alpha^2}{64s_w^3\pi^2} \frac{1}{M_W}
             \sum_{k=2}^3 q_{k1} \text{Im} \big( \rho_{\ell,11}^* q_{k2} \big)
             \big[ f_{8d}(x_{W h_k}) - f_{8d}(x_{W h_1}) \big] \\
                                & \quad + \frac{\alpha^2}{64s_w^3\pi^2} \frac{1}{M_W}
             \sum_{k=2}^3 q_{k1} \text{Im} \big( \rho_{\ell,11}^* q_{k2} \big)
             \big[ f_{8e}(x_{W h_k}, x_{W H^+}) - f_{8e}(x_{W h_1}, x_{W H^+}) \big] \,,
\end{split}
\end{equation}
where
\begin{align}
  f_{8a}(x,y) & = \log(x) \frac{1 + 2 x y}{y (x-1)} \,, \label{eq:f28a} \\
\begin{split} \label{eq:f28b}
  f_{8b}(x,y) & = \frac{12 y^2 - 16 y + 3}{1-x} \, x \, \Phi\bigg(\frac{1}{4y}\bigg)
                    + \log(x) \frac{10 x y + 1}{1-x} \\
& \quad 
+ \frac{12 x^2 y^2 - 16 x y + 3}{x-1} \Phi\bigg(\frac{1}{4xy}\bigg) \,,
\end{split}
\\
\begin{split} \label{eq:f28c}
f_{8c}(x) & = \big(12 x^2 - 16 x + 3 \big) \Phi\bigg(\frac{1}{4x}\bigg)
               - 2 \big[ \log(x) + 2 \big] \big(6 x + 1\big)
\end{split}
\\
\begin{split} \label{eq:f28d}
  f_{8d}(x) & = \frac{4 x^2 + 3 x}{9} \pi^2
+ \log^2(x) \frac{8 x^3 + 6 x^2 - 2}{6 x}
          - \log(x) \frac{8(x + 1)}{3}\\
& \quad 
          + \frac{8 x^4 + 6 x^3 - 2 x - 3}{3 x^2} \text{Li}_2(1 - x)\\
& \quad 
          + \frac{2 x^3 - 19 x^2 - 4 x + 3}{6 x^2} \Phi\bigg(\frac{1}{4x}\bigg)
          - \frac{8 x - 5}{3} \,,
\end{split}
\\
\begin{split} \label{eq:f28e}
f_{8e}(x,y) & = \log^2(x) \frac{x^3 y (3 - y)  
          + 3 x^2 y^2 - x (3y^2 + 4y +1) + y^2 + y}{2 x^3}\\
& \quad + \log(x) \log(y) \frac{x^3 y (y^2 - 4 y + 3) - 3 x^2 y^2 (y - 1)
                                + x y^2 (3y+1) - y^3}{2 x^3 (y - 1)}\\
& \quad 
+ \log(x) \frac{y - x y - x}{x^2}
+ \log(y) \frac{x^2 (1 + 7 y - 2 y^2) + xy (4y+1) - 2 y^2}{2 x^2 (y - 1)}\\
& \quad 
+ \bigg[ \frac{ x^4 y (3 - y^3 + 5 y^2 - 7 y) + x^3 y^3 ( 4 y - 8)}{2 x^4 (y - 1)}\\
& \quad \qquad
- \frac{x^2 y^2 (6 y^2 - y + 1) - x y^3 (4 y + 2) + y^4}{2 x^4 (y - 1)}
  \bigg] \Phi\bigg(y, \frac{y}{x}\bigg)\\
& \quad 
+ \frac{ x^3 y (3 - y) + 3 x^2 y^2 - x y (3 y + 4) + y (y + 1)}{x^3} \text{Li}_2(1 - x)\\
& \quad 
+ \frac{4 x^3 + 6 x^2 - 6 x + 1}{2 x^3 (y - 1)} \, y \, \Phi\bigg(\frac{1}{4x}\bigg)
+ \frac{x y (2 - x) + x - y}{x^2} \,.
\end{split}
\end{align}
The five terms in Eq.~\eqref{eq:eEDM:bosonic} correspond to
Barr-Zee-type diagrams with internal $Z$-boson exchange (first and
second lines), Barr-Zee-type diagrams with internal photon exchange
(third line), diagrams with internal $W$-boson exchange that are not
of the Barr-Zee type (fourth line), Barr-Zee-type diagrams with
internal $W$-boson exchange (fifth line). Note that this splitting is
not entirely physical, as it depends on the choice of gauge
parameter. Only in the sum of all contributions and after using the
orthogonality conditions of the Higgs mixing angles the gauge
parameters drops out, as we have verified explicitly. Note that the
first two terms in Eq.~\eqref{eq:eEDM:bosonic}, while suppressed by
the small $Z$-boson coupling to electrons, cannot be dropped as this
would lead to a gauge dependent result.

\begin{figure}[t]
  \centering
  \includegraphics[width=12em]{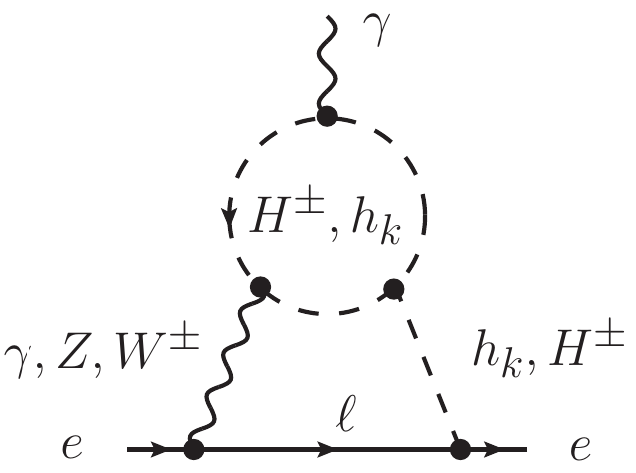}~~~
  \includegraphics[width=12em]{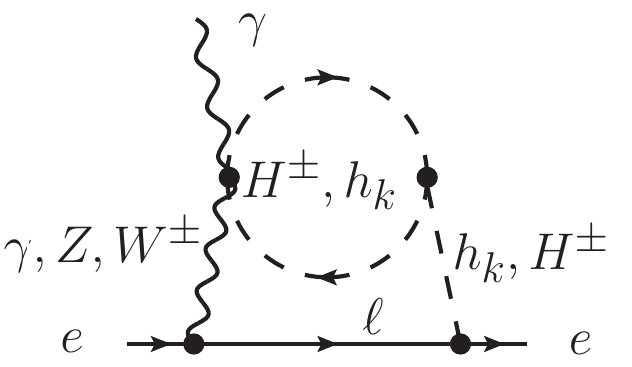}~~~
  \includegraphics[width=12em]{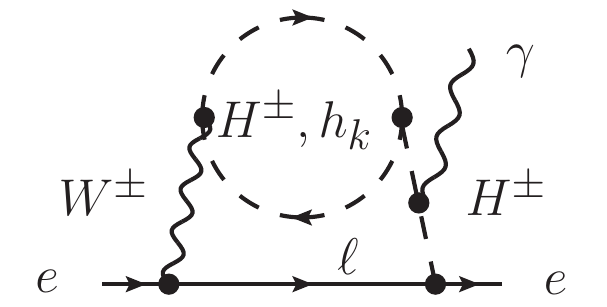}
  \caption{Representative two-loop diagrams contributing to the
    electron EDM with vertices arising from the Higgs potential. Here,
    $\ell$ denotes either an electron, $e$, or an electron neutrino,
    $\nu_e$. \label{fig:higgs:potential}}
\end{figure}

Last, we discuss the contributions of diagrams with vertices derived
from the Higgs potential. Such contribution necessarily enter at the
two-loop level (see Figure~\ref{fig:higgs:potential}). The
contribution of diagrams with internal photons and charged Higgs
bosons is
\begin{equation}
 d_e^{H^\pm \gamma} = \frac{\alpha e}{32\pi^3}
             \sum_{k=1}^3 \frac{v}{M_{h_k}^2} \text{Im} \big( \rho_{\ell,11}^* q_{k2} \big)
                  \Big[ q_{k1} Z_3 + \text{Re}(q_{k2} Z_7)
                  \Big] f_{9}(x_{H^\pm h_k}) \,,
\end{equation}
with the loop function
\begin{equation} \label{eq:f29}
  f_{9}(x) = x \, \Phi\bigg(\frac{1}{4x}\bigg) - \log(x) - 2 \,.
\end{equation}
The contribution of diagrams with internal $Z$ bosons and charged
Higgs bosons is
\begin{equation}
 d_e^{H^\pm Z} = \frac{\alpha e}{128\pi^3} \frac{c_w^2-s_w^2}{s_w c_w} v_e^Z
             \sum_{k=1}^3 \frac{v}{M_{h_k}^2} \text{Im} \big( \rho_{\ell,11}^* q_{k2} \big)
                  \Big[ q_{k1} Z_3 + \text{Re}(q_{k2} Z_7)
                  \Big] f_{10}(x_{H^\pm h_k}, x_{Z h_k}) \,,
\end{equation}
with
\begin{equation} \label{eq:f210}
  f_{10}(x,y) = \frac{1}{1 - y} \bigg[ \log(y) - x \, \Phi\bigg(\frac{1}{4x}\bigg)
                                       + \frac{x}{y} \Phi\bigg(\frac{y}{4x}\bigg) \bigg] \,.
\end{equation}
Note that these contributions are severely suppressed by the smallness
of $v_e^Z$, and are kept here only for completeness. The contribution
of diagrams with internal $W$ bosons and charged Higgs bosons is
\begin{equation}
 d_e^{H^\pm W} = \frac{\alpha e}{256s_w^2\pi^3}
             \sum_{k=1}^3 \frac{v}{M_{h_k}^2} \text{Im} \big( \rho_{\ell,11}^* q_{k2} \big)
                  \Big[ q_{k1} Z_3 + \text{Re}(q_{k2} Z_7)
                  \Big] f_{11}(x_{H^\pm h_k}, x_{W h_k}) \,,
\end{equation}
with
\begin{equation} \label{eq:f211}
\begin{split}
  f_{11}(x,y) & = 
\log(y) \log(x) \frac{y x - x^2 + 2 x - 1}{y^2(y-x)}
+ \log(y) \frac{2 x - y - 2}{y(y-x)} \\
& \quad + \log^2(x) \frac{2 x - 1}{x^2(x-y)}
+ \log(x) \frac{y x + 2 y - 2 x^2}{xy(y-x)}
+ \frac{2(x - 1)}{y x} \\
& \quad 
+ \frac{2 y x^2 - y^2 x  - y x - y - 
         x^3 + 3 x^2 - 3 x + 1}{y^2x(y-x)} \Phi\bigg(\frac{y}{x}, \frac{1}{x}\bigg) \\
& \quad 
+ \frac{2(y - 2 y x + x^3 - 2 x^2 + x)}{y^2 x^2} \text{Li}_2(1 - x)
+ \frac{2 x^2 - 4 x + 1}{x^2(x-y)} \Phi\bigg(\frac{1}{4x}\bigg) \,.
\end{split}
\end{equation}
The three contributions are separately gauge invariant. The
contribution of all diagrams with internal $Z$ bosons and vertices
from the Higgs potential that do not contain charged Higgs bosons are
zero.

All results presented in this section are new in the sense that, to
the best of our knowledge, they have never been calculated in fully
analytic form and for arbitrary phases in both the Higgs potential and
the Yukawa couplings. A more detailed comparison to previous results
can be found in Sec.~\ref{sec:conclusions}.

\boldmath
\section{Contributions to $\Gamma(\mu \to e \gamma)$} \label{sec:leg}
\unboldmath

In this section, we present the leading contributions to the decay
rate of the radiative decay $\mu \to e \gamma$. (The generalization to
tau decay is discussed in Sec.~\ref{sec:tau:decay}.) The diagrams to
be evaluated for this process are very similar to those discussed in
the previous section, which allows for valuable cross checks of our
results. We parameterize the dipole contributions to the
$\mu \to e \gamma$ decay by using the effective Lagrangian
\begin{equation} \label{L_eff_muea_LR}
\begin{split}
\mathcal L_\text{eff} & =
 - c_L \frac{e m_\mu}{16\pi^2 v^2}
     (\bar e \sigma^{\rho\sigma} P_L \mu) F_{\rho\sigma}
 - c_R \frac{e m_\mu}{16\pi^2 v^2}
     (\bar e \sigma^{\rho\sigma} P_R \mu) F_{\rho\sigma}
 + \text{h.c.} \,.
\end{split}
\end{equation}
Here, $P_R = (1+\gamma_5)/2$ and $P_L = (1-\gamma_5)/2$ are the left
and right chiral projectors, respectively. In analogy to our strategy
in the previous section, we neglect all higher-order effects (in
particular, the effects of QED running), such that this Lagrangian can
be viewed as valid at scales of the order of the muon mass. Once the
coefficients $c_L$ and $c_R$ have been determined by a matching
calculation at the weak scale, the corresponding total decay rate is
obtained as
\begin{equation}
\begin{split}
  \Gamma(\mu \to e + \gamma)
  = \frac{m_\mu^5 \alpha}{256 \pi^4 v^4}
    \Big( |c_R|^2 + |c_L|^2 \Big) \,.
\end{split}
\end{equation}

Again, we present our results for the dipole coefficients by splitting
them according to the number of loops, defining
\begin{equation}
  c_{L/R} = c_{L/R}^\text{one-loop} + c_{L/R}^\text{two-loop} + \ldots \,.
\end{equation}
We find the following one-loop contributions:
\begin{align}
\begin{split}
  c_R^\text{one-loop}
& = \sum_{k} \frac{v^2}{4M_{h_k}^2}
           \bigg[  \sum_{i=2,3} \frac{m_{\ell_i}}{m_{\mu}}
                   \rho_{\ell,2i}^* \rho_{\ell,i1}^* \big(q_{k2} \big)^2
                   \bigg( 3 + 2 \log \frac{m_{\ell_i}^2}{M_{h_k}^2} \bigg)
                 - \frac{1}{3} \sum_{i=1}^3
                   \rho_{\ell,i2} \rho_{\ell,i1}^* |q_{k2}|^2 \\
& \hspace{6em} + \frac{m_{\mu}}{v} 
                   \rho_{\ell,21}^* q_{k2} q_{k1}
                   \bigg( \frac{8}{3} + 2 \log \frac{m_{\mu}^2}{M_{h_k}^2} \bigg)
                   \bigg] \,,
\end{split}\\
\begin{split}
  c_L^\text{one-loop}
& = \sum_{k} \frac{v^2}{4M_{h_k}^2}
           \bigg[  \sum_{i=2,3} \frac{m_{\ell_i}}{m_{\mu}}
                   \rho_{\ell,1i} \rho_{\ell,i2} \big(q_{k2}^* \big)^2
                   \bigg( 3 + 2 \log \frac{m_{\ell_i}^2}{M_{h_k}^2} \bigg)
                 - \frac{1}{3} \sum_{i=1}^3
                   \rho_{\ell,2i}^* \rho_{\ell,1i} |q_{k2}|^2 \\
& \hspace{6em} + \frac{m_{\mu}}{v} 
                   \rho_{\ell,12} q_{k2}^* q_{k1}
                   \bigg( \frac{8}{3} + 2 \log \frac{m_{\mu}^2}{M_{h_k}^2} \bigg)
                 \bigg]
               + \frac{v^2}{12M_{H^\pm}^2} \sum_{i=1}^3 \rho_{\ell,2i}^* \rho_{\ell,1i} \,,
\end{split}
\end{align}
where we have dropped terms that are suppressed by a factor of
$m_e/m_\mu$ or $m_e/m_\tau$.

Also the $\mu \to e \gamma$ decay rate can receive numerically large
contributions from two-loop diagrams (see
Ref.~\cite{Bjorken:1977vt}). As before, we calculate all leading terms
that are quadratic in the Yukawa couplings and are not mere
corrections to coupling combinations that appear already at
one-loop. We split the two-loop contribution into several terms that
are separately gauge invariant,
\begin{equation}
  c_{R,L}^\text{two-loop}
= \sum_{f_i = t,b,c,\tau} c_{R,L}^{f_i h}
  + \sum_{f_i = t,c,\tau} c_{R,L}^{f_i H^\pm}
  + \big( c_{R,L}^{hZ} + c_{R,L}^\text{kin.} \big)
  + \big( c_{R,L}^{H^\pm \gamma} + c_{R,L}^{H^\pm Z} + c_{R,L}^{H^\pm W} \big) \,,
\end{equation}
where the individual terms receive contributions from the same type of
diagrams as discussed for the electron EDM. The explicit results are
given in the following. The two-loop diagrams with heavy-fermion loops
and photon or $Z$-boson exchange give a contribution
\begin{align}
\begin{split}\label{eq:cR:ht}
  c_R^{f_i h}
& = \frac{3\alpha Q_{f_i}}{4\pi} \frac{m_{f_i}}{m_\mu}
    \sum_{k=1}^3 \frac{v^2}{M_{h_k}^2} \\
& \times \bigg\{ 
  \left( Q_{f_i} \big[f_{2} + f_{1}\big](x_{f_i h_k})
         + \frac{v_{f_i}^Z v_e^Z}{4} \big[f_{4} + f_{3}\big](x_{f_i h_k}, x_{f_i Z}) \right)
           \rho_{\ell,21}^* \rho_{f,ii} |q_{k2}|^2 \\
& \qquad + \left( Q_{f_i} \big[f_{2} - f_{1}\big](x_{f_i h_k})
                 + \frac{v_{f_i}^Z v_e^Z}{4}
                   \big[f_{4} - f_{3}\big](x_{f_i h_k}, x_{f_i Z}) \right)
           \rho_{\ell,21}^* \rho_{f,ii}^* \big(q_{k2}\big)^2 \\
& \qquad + 2 \left( Q_{f_i} f_{2}(x_{f_i h_k})
                 + \frac{v_{f_i}^Z v_e^Z}{4} f_{4}(x_{f_i h_k}, x_{f_i Z}) \right)
           \frac{m_{f_i}}{v} \rho_{\ell,21}^* q_{k2} q_{k1}
  \bigg\} \,,
\end{split}
\end{align}
for $f_i = t,c$, and
\begin{align}
\begin{split}\label{eq:cR:hb}
  c_R^{f_i h}
& = \frac{N_c(f_i)\alpha Q_{f_i}}{4\pi} \frac{m_{f_i}}{m_\mu}
    \sum_{k=1}^3 \frac{v^2}{M_{h_k}^2} \\
& \times \bigg\{ 
  \left( Q_{f_i} \big[f_{2} + f_{1}\big](x_{f_i h_k})
         + \frac{v_{f_i}^Z v_e^Z}{4} \big[f_{4} + f_{3}\big](x_{f_i h_k}, x_{f_i Z}) \right)
           \rho_{\ell,21}^* \rho_{f,ii}^* \big(q_{k2}\big)^2 \\
& \qquad + \left( Q_{f_i} \big[f_{2} - f_{1}\big](x_{f_i h_k})
                 + \frac{v_{f_i}^Z v_e^Z}{4}
                   \big[f_{4} - f_{3}\big](x_{f_i h_k}, x_{f_i Z}) \right)
           \rho_{\ell,21}^* \rho_{f,ii} |q_{k2}|^2 \\
& \qquad + 2 \left( Q_{f_i} f_{2}(x_{f_i h_k})
                 + \frac{v_{f_i}^Z v_e^Z}{4} f_{4}(x_{f_i h_k}, x_{f_i Z}) \right)
                 \frac{m_{f_i}}{v} \rho_{\ell,21}^* q_{k2} q_{k1}
  \bigg\} \,,
\end{split}
\end{align}
for $f_i = b,\tau$. The results for $c_L^{f_i}$ can be obtained from
these expressions by simply replacing $\rho_{f,ij}^* \to \rho_{f,ji}$
and $q_{k2} \to q_{k2}^*$. Here,
$[f_{2} \pm f_{1}](x) \equiv f_{2}(x) \pm f_{1}(x)$, and
$[f_{3} \pm f_{4}](x,y) \equiv f_{4}(x,y) \pm f_{3}(x,y)$. The loop
functions have been defined above, in
Eqs.~\eqref{eq:f21}-\eqref{eq:f24} (their limiting behavior for small
mass ratios is given in
Eqs.~\eqref{eq:limit:f21}-\eqref{eq:limit:f24}).

The contribution of diagrams with virtual charged Higgs bosons and top
quarks is
\begin{align}
  c_R^{tH^\pm} & = \frac{\alpha}{8\pi s_w^2} \frac{m_{t}}{m_\mu} \frac{v^2}{M_{H^\pm}^2}
                \bigg[
                \rho_{u,33} \rho_{\ell,21}^*
                f_{5}(x_{t H^\pm}, x_{t W})
                + \frac{m_b}{m_t} \rho_{d,33} \rho_{\ell,21}^*
                f_{6}(x_{t H^\pm}, x_{t W})
                \bigg] \,,
\end{align}
the contribution of diagrams with internal charm quarks is
\begin{align}\label{eq:cRcH}
\begin{split}
  c_R^{cH^\pm}
= & \frac{\alpha}{16\pi s_w^2} \frac{m_c}{m_\mu} \frac{v^2}{M_{H^\pm}^2}
    \rho_{u,22} \rho_{\ell,21}^*
                 \bigg[ 13 + 4 \log\bigg(\frac{m_c^2}{M_{H^\pm}^2}\bigg)
                           + 4 \log\bigg(\frac{m_c^2}{M_W^2}\bigg)
                 \bigg]
    \frac{\log x_{W H^\pm}}{1 - x_{W H^\pm}} \,,
\end{split}
\end{align}
and the contribution of diagrams with charged Higgs bosons and tau
leptons is
\begin{align}
  c_R^{\tau H^\pm} & = \frac{\alpha}{16\pi s_w^2} \frac{v^2}{M_{H\pm}^2} \frac{m_{\tau}}{m_\mu}
             \rho_{\ell,33} \rho_{\ell,21}^*
             \frac{1}{(1 - x_{W H^\pm})} \log x_{W H^\pm} \,.
\end{align}
The coefficients $c_L^{t/c/\tau H^\pm}$ can be obtained from these
results by replacing $\rho_{f,ij}^* \to \rho_{f,ji}$.

The contribution of diagrams with vertices arising from the kinetic
terms of the Higgs bosons can again be split into two groups that are
separately gauge invariant. The first are diagrams with internal $Z$
bosons and neutral Higgs bosons that are not of the
``Bjorken-Weinberg'' type. They give
\begin{align}
\begin{split}
  c_R^{h Z}
 & = - \frac{e\alpha \big(v_e^Z\big)^2}{48\pi s_w c_w} \frac{v^2}{M_Z m_\mu}
             \sum_{k=2}^3 q_{k1} q_{k2} \rho_{\ell,21}^*
             \big[ f_{7+}(x_{Z h_k}) - f_{7+}(x_{Z h_1}) \big] \\
 & \quad - \frac{e\alpha \big(a_e^Z\big)^2}{48\pi s_w c_w} \frac{v^2}{M_Z m_\mu}
             \sum_{k=2}^3 q_{k1}  q_{k2} \rho_{\ell,21}^*
             \big[ f_{7-}(x_{Z h_k}) - f_{7-}(x_{Z h_1}) \big] \,,
\end{split}
\end{align}
with the loop functions $f_{7\pm}(x)$ defined in
Eq.~\eqref{eq:f27pm}. The coefficient $c_L^{h Z}$ can be obtained from
$c_R^{h Z}$ by the replacements
$\rho_{f,ij} \leftrightarrow \rho_{f,ji}^*$ and
$q_{k2} \leftrightarrow q_{k2}^*$. All remaining two-loop diagrams
involving vertices from the Higgs kinetic terms give
\begin{align}
\begin{split}
  c_R^{h,\text{kin.}} & = \quad \frac{e\alpha v_e^Z}{32\pi} \frac{v^2}{M_Z m_\mu}
             \sum_{k=2}^3 q_{k1} \rho_{\ell,21}^* q_{k2}
             \big[ f_{8a}(x_{Z h_k}, c_w^2) - f_{8a}(x_{Z h_1}, c_w^2) \big] \\
                           & \quad + \frac{e\alpha v_e^Z}{32\pi s_w^2} \frac{v^2}{M_Z m_\mu}
             \sum_{k=2}^3 q_{k1} \rho_{\ell,21}^* q_{k2}
             \big[ f_{8b}(x_{Z h_k}, c_w^2) - f_{8b}(x_{Z h_1}, c_w^2) \big] \\
                           & \quad + \frac{e\alpha}{16s_w\pi} \frac{v^2}{M_W m_\mu}
             \sum_{k=2}^3 q_{k1} \rho_{\ell,21}^* q_{k2}
             \big[ f_{8c}(x_{W h_k}) - f_{8c}(x_{W h_1}) \big] \\
                           & \quad + \frac{e\alpha}{32s_w^3\pi} \frac{v^2}{M_W m_\mu}
             \sum_{k=2}^3 q_{k1} \rho_{\ell,21}^* q_{k2}
             \big[ f_{8d}(x_{W h_k}) - f_{8d}(x_{W h_1}) \big] \\
                           & \quad + \frac{e\alpha}{32s_w^3\pi} \frac{v^2}{M_W m_\mu}
             \sum_{k=2}^3 q_{k1} \rho_{\ell,21}^* q_{k2}
             \big[ f_{8e}(x_{W h_k}, x_{W H^+}) - f_{8e}(x_{W h_1}, x_{W H^+}) \big] \,,
\end{split}
\end{align}
where the loop functions $f_{8i}$, $i=a,b,c,d,e$, are defined in
Eqs.~\eqref{eq:f28a}-\eqref{eq:f28e}. Again, $c_L^{h,\text{kin.}}$ can
be obtained from $c_R^{h,\text{kin.}}$ by the replacements
$\rho_{f,ij}^* \to \rho_{f,ji}$ and $q_{k2} \to q_{k2}^*$.

Finally, the contribution of diagrams with vertices from the Higgs
potential and internal photons and charged Higgs is
\begin{align}
 c_R^{H^\pm \gamma} & = \frac{\alpha}{4\pi} \frac{v}{m_\mu}
             \sum_{k=1}^3 \frac{v^2}{M_{h_k}^2}
             \rho_{\ell,21}^* q_{k2}
   \bigg( q_{k1} Z_3 + \text{Re}(q_{k2} Z_7)
                  \bigg) f_{9}(x_{H^\pm h_k}) \,, \\
 c_L^{H^\pm \gamma} & = \frac{\alpha}{4\pi} \frac{v}{m_\mu}
             \sum_{k=1}^3 \frac{v^2}{M_{h_k}^2}
             \rho_{\ell,12} q_{k2}^*
   \bigg( q_{k1} Z_3 + \text{Re}(q_{k2} Z_7)
                  \bigg) f_{9}(x_{H^\pm h_k}) \,,
\end{align}
with $f_{9}(x)$ defined in Eq.~\eqref{eq:f29}. The contribution of
diagrams with vertices from the Higgs potential and $Z$ bosons and
charged Higgs bosons is
\begin{align}
 c_R^{H^\pm Z} = \frac{\alpha}{16\pi} \frac{c_w^2-s_w^2}{s_w c_w} v_e^Z
              \frac{v}{m_\mu} \sum_{k=1}^3 \frac{v^2}{M_{h_k}^2}
              \rho_{\ell,21}^* q_{k2}
                  \bigg( q_{k1} Z_3 + \text{Re}(q_{k2} Z_7)
                  \bigg) f_{10}(x_{H^\pm h_k}, x_{Z h_k}) \,, \\
 c_L^{H^\pm Z} = \frac{\alpha}{16\pi} \frac{c_w^2-s_w^2}{s_w c_w} v_e^Z
             \frac{v}{m_\mu} \sum_{k=1}^3 \frac{v^2}{M_{h_k}^2}
             \rho_{\ell,12} q_{k2}^*
                  \bigg( q_{k1} Z_3 + \text{Re}(q_{k2} Z_7)
                  \bigg) f_{10}(x_{H^\pm h_k}, x_{Z h_k}) \,,
\end{align}
with $f_{10}$ defined in Eq.~\eqref{eq:f210}. The contribution of
diagrams with vertices from the Higgs potential and internal $W$
bosons and charged Higgs bosons is
\begin{align}
 c_R^{H^\pm W} = \frac{\alpha}{32s_w^2\pi}
             \frac{v}{m_\mu} \sum_{k=1}^3 \frac{v^2}{M_{h_k}^2}
             \rho_{\ell,21}^* q_{k2}
                  \bigg( q_{k1} Z_3 + \text{Re}(q_{k2} Z_7)
                  \bigg) f_{11}(x_{H^\pm h_k}, x_{W h_k}) \,, \\
 c_L^{H^\pm W} = \frac{\alpha}{32s_w^2\pi}
             \frac{v}{m_\mu} \sum_{k=1}^3 \frac{v^2}{M_{h_k}^2}
             \rho_{\ell,12} q_{k2}^*
                  \bigg( q_{k1} Z_3 + \text{Re}(q_{k2} Z_7)
                  \bigg) f_{11}(x_{H^\pm h_k}, x_{W h_k}) \,,
\end{align}
with $f_{11}$ defined in Eq.~\eqref{eq:f211}. As for the electron EDM,
there are no diagrams with vertices from the Higgs potential with
internal photons and neutral Higgs bosons, while the contribution from
diagrams with internal $Z$ bosons and neutral Higgs bosons is zero.

\boldmath
\subsection{$\tau \to \ell \gamma$ decay}\label{sec:tau:decay}
\unboldmath

Defining the effective Lagrangian relevant for
$\tau \to \ell_j \gamma$ decay, with $j=1,2$, as
\begin{equation} \label{L_eff_tau_LR}
\begin{split}
\mathcal L_\text{eff} & =
 - \sum_{j=1,2} c_{j,L}^{\tau} \frac{e m_\tau}{16\pi^2 v^2}
     (\bar \ell_j \sigma^{\rho\sigma} P_L \tau) F_{\rho\sigma}
 - \sum_{j=1,2} c_{j,R}^{\tau} \frac{e m_\tau}{16\pi^2 v^2}
     (\bar \ell_j \sigma^{\rho\sigma} P_R \tau) F_{\rho\sigma}
 + \text{h.c.} \,,
\end{split}
\end{equation}
the corresponding total decay rate is
\begin{equation}
\begin{split}
  \Gamma(\tau \to \ell_j + \gamma)
  = \frac{m_\tau^5 \alpha}{256 \pi^4 v^4}
    \Big( |c_{j,R}^\tau|^2 + |c_{j,L}^\tau|^2 \Big) \,.
\end{split}
\end{equation}
Then the one-loop results are given by
\begin{align}
\begin{split}
  c_{j,R}^{\tau,\text{one-loop}}
& = \sum_{k} \frac{v^2}{4M_{h_k}^2}
           \bigg[  \sum_{i=2,3} \frac{m_{\ell_i}}{m_{\tau}}
                   \rho_{\ell,3i}^* \rho_{\ell,ij}^* \big(q_{k2} \big)^2
                   \bigg( 3 + 2 \log \frac{m_{\ell_i}^2}{M_{h_k}^2} \bigg)
                 - \frac{1}{3} \sum_{i=1}^3
                   \rho_{\ell,i3} \rho_{\ell,ij}^* |q_{k2}|^2 \\
& \hspace{6em} + \frac{m_{\tau}}{v} 
                   \rho_{\ell,3j}^* q_{k2} q_{k1}
                   \bigg( \frac{8}{3} + 2 \log \frac{m_{\tau}^2}{M_{h_k}^2} \bigg)
                   \bigg] \,,
\end{split}\\
\begin{split}
  c_{j,L}^{\tau,\text{one-loop}}
& = \sum_{k} \frac{v^2}{4M_{h_k}^2}
           \bigg[  \sum_{i=2,3}  \frac{m_{\ell_i}}{m_{\tau}}
                   \rho_{\ell,ji} \rho_{\ell,i3} \big(q_{k2}^* \big)^2
                   \bigg( 3 + 2 \log \frac{m_{\ell_i}^2}{M_{h_k}^2} \bigg)
                 - \frac{1}{3} \sum_{i=1}^3
                   \rho_{\ell,3i}^* \rho_{\ell,ji} |q_{k2}|^2 \\
& \hspace{6em} + \frac{m_{\tau}}{v} 
                   \rho_{\ell,j3} q_{k2}^* q_{k1}
                   \bigg( \frac{8}{3} + 2 \log \frac{m_{\tau}^2}{M_{h_k}^2} \bigg)
                   \bigg]
               + \frac{v^2}{12M_{H^\pm}^2} \sum_{i=1}^3 \rho_{\ell,ji} \rho_{\ell,3i}^* \,,
\end{split}
\end{align}
while for the two-loop contributions, in $c_{R/L}^{\text{two-loop}}$
replace $1/m_\mu$ by $1/m_\tau$, $\rho_{\ell,12}$ by $\rho_{\ell,j3}$,
and $\rho_{\ell,21}^*$ by $\rho_{\ell,3j}^*$ to obtain
$c_{j,R/L}^{\tau,\text{two-loop}}$.

\section{Discussion} \label{sec:conclusions}

We presented the leading one-loop and two-loop contributions to the
electron EDM and the radiative lepton flavor violating decays
$\mu \to e + \gamma$ and $\tau \to e/\mu + \gamma$ in the
unconstrained 2HDM, keeping terms that are quadratic in the Yukawa
interactions. To the best of our knowledge, these results have been
presented here in full generality for the first time.

We verified our results through several consistency checks. We
confirmed that all infrared and ultraviolat divergences cancel in our
results. Moreover, we performed the calculation in generalized $R_\xi$
gauge for all gauge bosons, and verified explicitly that all our
results are gauge parameter independent. Moreover, we were able to
reproduce a number of results in the literature using our
expressions. In particular, we can check part of our results against
the expressions presented in Ref.~\cite{Altmannshofer:2020shb}, by
taking the limit of real and diagonal Yukawa couplings. To make
contact with the complex 2HDM that has been discussed in
Ref.~\cite{Altmannshofer:2020shb}, we define the coefficients
\begin{equation}
 c_{u,ij} = - \frac{\rho_{u,ij} v}{m_{u_i}} \,, \qquad
 c_{d,ij} = \frac{\rho_{d,ij} v}{m_{d_i}} \,, \qquad
 c_{\ell,ij} = \frac{\rho_{\ell,ij} v}{m_{\ell_i}} \,.
\end{equation}
The expressions in Ref.~\cite{Altmannshofer:2020shb} can then be
obtained by choosing the coupling matrices to be real and diagonal,
i.e., $c_{f,ij} = c_{f} \delta_{ij}$ with real and horizontally
flavor-universal coefficients $c_f$. We find that our results
reproduce those in Ref.~\cite{Altmannshofer:2020shb} in this limit,
with the exception of the contributions $d_e^{H^\pm \gamma}$,
$d_e^{H^\pm Z}$, and $d_e^{H^\pm W}$, which are absent in the complex
2HDM and are presented here for the first time. As a further check, we
verified that our one-loop results for $\mu \to e \gamma$ are
consistent with the expressions presented in
Ref.~\cite{Hisano:1995cp}, after adjusting the couplings appropriately
and taking the limit of small fermion masses. Finally, by taking
appropriate linear combinations of our results for $\mu \to e \gamma$,
and replacing the Higgs couplings to muons with the corresponding
couplings to electrons, we reproduce all two-loop contributions to the
electron EDM. (The one-loop contributions cannot be simply reproduced
in that way, since the equations of motion contribute differently in
the two cases.)

Some two-loop results for $\mu \to e \gamma$ involving the exchange of
virtual neutral Higgs bosons have been presented in
Ref.~\cite{Chang:1993kw}. We have not attempted a detailed comparison,
since their results are partially given in numerical form (using
results of Ref.~\cite{Leigh:1990kf}), and partially in terms of
parametric integrals. Moreover, our results show that the bosonic
contributions without the inclusion of charged Higgs bosons are gauge
dependent.

\subsection*{Python code}

For the convenience of the reader, we provide an implementation of our
results into \texttt{python} code. It can be downloaded via the public
git repository
\begin{center}
\url{https://gitlab.com/jbrod/general-2hdm-pheno}
\end{center}
and provides numerical values for all Wilson coefficients and decay
rates, for user-specified model input parameters. Further information
on the usage can also be found in the repository.

\section*{Acknowledgments}

We thank Stefania Gori and Reinard Primulando for helpful discussions,
and Duncan Rocha for pointing out a mistake in
equations~\eqref{eq:cR:ht} and~\eqref{eq:cR:hb} in earlier versions of
this paper.
The research of W.A. is supported by the U.S. Department of Energy
grant number DE-SC0010107.
J.B. thanks Emmanuel Stamou for many discussion and comments on the
manuscript, and Emmanuel Stamou and Tom Steudtner for consistency
checks of some \texttt{MaRTIn} routines. J.B. acknowledges support by
DoE grant DE-SC0011784.
P.U. thanks the High-Energy Physics Research Unit, Chulalongkorn
University for the hospitality while part of this work is being
completed, and acknowledges support from the Mid-Career Research Grant
from the National Research Council of Thailand under contract
no.~N42A650378.
%
The Feynman diagrams in the figures have been generated using
\texttt{jaxodraw}~\cite{Binosi:2003yf}, based on
\texttt{axodraw}~\cite{Vermaseren:1994je}.

\appendix

\section{Parameters of the Higgs Potential} \label{app:higgs_potential}

In this appendix, we collect some useful equations regarding the
scalar potential of the 2HDM~\cite{Davidson:2005cw}. First, the
conditions of minimization of the Higgs potential given in
Eq.~\eqref{eq:genpot}, namely,
\begin{align}
\begin{split}
m_{11}^2 &= {\rm{Re}}(m_{12}^2\,e^{i\zeta})\,\frac{v_2}{v_1} \\
& \quad  -\frac{1}{2}
\left[\lambda_1v_1^2+\lambda_{345}\,v_2^2 + 
 v_1v_2{\rm{Re}}(2\lambda_6e^{i\zeta}+\lambda_6^*e^{-i\zeta})+\frac{v_2^3}{v_1}{\rm{Re}}(\lambda_7e^{i\zeta})\right] \,,
\label{minconditionsa}
\end{split} \\
\begin{split}
m_{22}^2 &= {\rm{Re}}(m_{12}^2\,e^{i\zeta})\,\frac{v_1}{v_2} \\
& \quad -\frac{1}{2}
\left[\lambda_2 v_2^2+\lambda_{345}\,v_1^2
+ \frac{v_1^3}{v_2}{\rm{Re}}(\lambda_6^*e^{-i\zeta})+v_1v_2{\rm{Re}}(\lambda_7e^{i\zeta}+2\lambda_7^*e^{-i\zeta})\right] \,,
\end{split} \\
 {\rm{Im}}(m_{12}^2 e^{i\zeta}) &= \frac{1}{2}\left(v_1v_2{\rm{Im}}(\lambda_5 e^{2i\zeta})+v_1^2{\rm{Im}}(\lambda_6 e^{i\zeta})+
 v_2^2{\rm{Im}}(\lambda_7 e^{i\zeta})\right) \,,
\label{minconditionsb}
\end{align}
can be used to determine $v_1$ and $v_2$. Here, we have defined
$\lambda_{345} = \lambda_3 + \lambda_4 + \text{Re}(\lambda_5
e^{2i\zeta})$.  The Higgs potential can also be expressed in the Higgs
basis defined in Eq.~\eqref{eq:HBpot}. The corresponding mass terms
and quartic interactions are linearly related to the $\lambda_i$,
$m_{ij}^2$:
\begin{align}
Y_1 & = m_{11}^2 c_\beta^2 +m_{22}^2 s_\beta^2-{\rm{Re}}(m_{12}^2 e^{i\zeta})s_{2\beta} \,, \\
Y_2 & = m_{11}^2 s_\beta^2 +m_{22}^2 c_\beta^2+{\rm{Re}}(m_{12}^2 e^{i\zeta})s_{2\beta} \,, \\
Y_3 e^{i\zeta} & = \tfrac{1}{2}(m_{11}^2-m_{22}^2)s_{2\beta}+{\rm{Re}}(m_{12}^2 e^{i\zeta})c_{2\beta}
                  +i\,{\rm{Im}}(m_{12}^2 e^{i\zeta}) \,, \\
    \label{eq:Zlist}
Z_1 & = \lambda_1 c^4_\beta +\lambda_2 s^4_\beta +\textstyle\frac{1}{2}\lambda_{345}s_{2\beta}
 +2s_{2\beta}\left[c_\beta^2{\rm{Re}}(\lambda_6 e^{i\zeta})+s_\beta^2{\rm{Re}}(\lambda_7 e^{i\zeta})\right]  \,, \\
Z_2 & = \lambda_1s^4_\beta +\lambda_2 c^4_\beta + \textstyle\frac{1}{2}\lambda_{345}s_{2\beta}^2
       -2s_{2\beta}\left[s_\beta^2{\rm{Re}}(\lambda_6 e^{i\zeta})+c_\beta^2{\rm{Re}}(\lambda_7 e^{i\zeta})\right]  \,, \\
Z_3 & = \tfrac{1}{4}s_{2\beta}^2\left(\lambda_1+\lambda_2-2\lambda_{345}\right)+\lambda_3
     -s_{2\beta}c_{2\beta}{\rm{Re}}((\lambda_6-\lambda_7)e^{i\zeta}) \,, \\
Z_4 & = \tfrac{1}{4}s_{2\beta}^2\left(\lambda_1+\lambda_2-2\lambda_{345}\right)+\lambda_4
       -s_{2\beta}c_{2\beta}{\rm{Re}}((\lambda_6-\lambda_7)e^{i\zeta}) \,, \\
\begin{split}
Z_5 e^{2i\zeta} & = \tfrac{1}{4}s_{2\beta}^2\left(\lambda_1+\lambda_2-2\lambda_{345}\right)+{\rm{Re}}(\lambda_5 e^{2i\zeta}) +i c_{2\beta}{\rm{Im}}(\lambda_5 e^{2i\zeta}) \\
 & \quad -s_{2\beta}c_{2\beta}{\rm{Re}}((\lambda_6-\lambda_7)e^{i\zeta})
         -is_{2\beta}{\rm{Im}}((\lambda_6-\lambda_7)e^{i\zeta})
\end{split} \,, \\
\begin{split}
Z_6 e^{i\zeta} & = -\tfrac{1}{2}s_{2\beta}\big(\lambda_1 c_\beta^2 -\lambda_2 s_\beta^2-\lambda_{345} c_{2\beta} -i{\rm{Im}}(\lambda_5 e^{2i\zeta})\big) \\
& \quad +c_\beta c_{3\beta}{\rm{Re}}(\lambda_6e^{i\zeta})+s_\beta s_{3\beta}{\rm{Re}}(\lambda_7e^{i\zeta})
 +ic_\beta^2{\rm{Im}}(\lambda_6e^{i\zeta})+is_\beta^2{\rm{Im}}(\lambda_7e^{i\zeta})
\end{split} \,, \\
\begin{split}
  Z_7 e^{i\zeta} & = -\tfrac{1}{2} s_{2\beta}\big(\lambda_1 s_\beta^2 -\lambda_2 c_\beta^2 +\lambda_{345} c_{2\beta}+i{\rm{Im}}(\lambda_5 e^{2i\zeta})\big) \\
& \quad +s_\beta s_{3\beta}{\rm{Re}}(\lambda_6e^{i\zeta})+c_\beta c_{3\beta}{\rm{Re}}(\lambda_7e^{i\zeta})
 +is_\beta^2{\rm{Im}}(\lambda_6e^{i\zeta})+ic_\beta^2{\rm{Im}}(\lambda_7e^{i\zeta}) \,.
\end{split}
\end{align}

\section{Calculation of fermion-loop contributions in the HV scheme}

Some of the Barr-Zee diagrams contain closed fermion loops (see
Fig.~\ref{fig:BZ:fermion}), leading to traces of products of Dirac
matrices including $\gamma_5$ matrices. The $\gamma_5$ matrices arise
from pseudo-scalar or axial couplings of the Higgs and gauge
bosons. It is well-known that fully anti-commuting $\gamma_5$, i.e.,
$\{\gamma^\mu, \gamma_5\} \equiv \gamma^\mu \gamma_5 - \gamma_5
\gamma^\mu = 0$ for all $\mu$ (the ``NDR'' prescription) is
algebraically inconsistent in these cases~\cite{Collins:1984xc}. To
the best of our knowlegde, this issue has never been discussed in the
context of Barr-Zee diagrams, and the NDR prescription has always been
used in the literature. In this work, we have employed the so-called
``HV'' scheme~\cite{tHooft:1972tcz, Breitenlohner:1977hr} whenever a
closed fermion loop is present. The results obtained in this
consistent scheme are in agreement with the know results in the
literature, thus providing a valuable check on previous
calculations. In the remainder of this section, we provide some of the
details of our calculation.

\begin{figure}[t]
  \centering
  \includegraphics[height=8em]{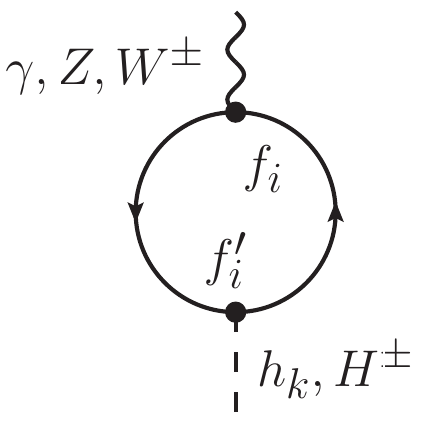}~~~~~~~
  \includegraphics[height=8em]{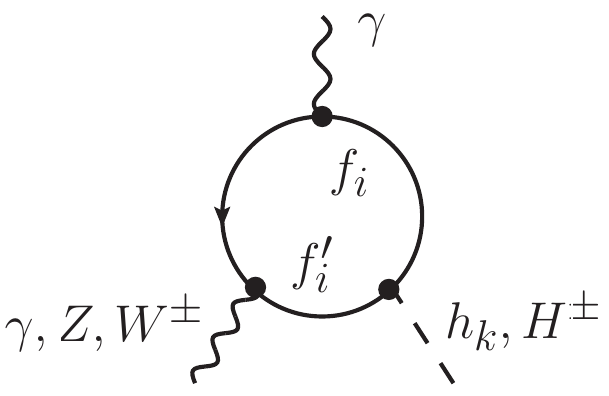}
  \caption{Counterterm insertions whose divergent parts contribute to
    the electron EDM in the HV scheme. \label{fig:HV:count:ins}}
\end{figure}

The `t~Hooft-Veltman (HV) prescription for $\gamma_5$ in
$d = 4 - 2\epsilon$ spacetime dimensions is~\cite{Collins:1984xc}
$\{\gamma^\mu, \gamma_5\} = 0$ for $\mu=0,1,2,3$, and
$\gamma^\mu \gamma_5 = \gamma_5 \gamma^\mu$ otherwise. These mixed
(anti-)commutation relations have been implemented in the
\texttt{MaRTIn} package.
\begin{figure}[t]
  \centering
  \includegraphics[width=12em]{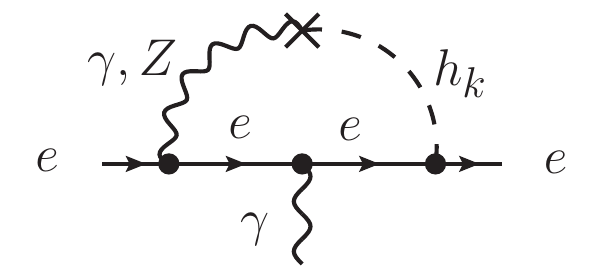}~~~
  \includegraphics[width=12em]{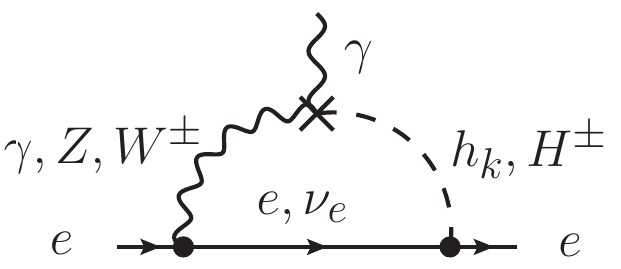}
  \caption{Counterterm diagrams that contribute to the electron EDM in
    the HV scheme. \label{fig:HV:count}}
\end{figure}
The calculation of the fermionic Barr-Zee contributions is performed
as follows. We calculate the two-loop contributions in the HV scheme,
dropping all terms that vanish in the limit $d \to 4$. The crucial
point is that there are also one-loop contributions, arising from the
two counterterm diagrams displayed in Fig.~\ref{fig:HV:count}. (These
diagrams vanish when calculated using the NDR
prescription, while in the HV scheme they give contributions of order
$\epsilon$.) We then also need to calculate the counterterm insertions
at one-loop (see Fig.~\ref{fig:HV:count:ins}). Since we work in the
$\overline{\text{MS}}$ scheme, only the divergent part of these
diagrams is needed.\footnote{We checked explicitly that no finite
  contribution to the counterterm insertion leads to a non-vanishing
  contribution to the results in the limit $d \to 4$.} In these
calculations it is of crucial importance to keep all ``evanescent''
contributions (i.e., terms of order $\epsilon$ as well as terms
proportional to components of momenta in the ``parallel space'',
$\mu > 3$) in intermediate stages of the calculation. After inclusion
of all these terms we recover the results that have been obtained
previously using the NDR prescription.

\addcontentsline{toc}{section}{References}
\bibliographystyle{JHEP}
\bibliography{references}{}

\end{document}